\documentclass[10pt,letterpaper,twocolumn]{article}
\usepackage[T1]{fontenc}
\usepackage[utf8]{inputenc}
\usepackage{libertine}
\usepackage{microtype}
\usepackage[left=0.7in,right=0.7in,top=0.65in,bottom=0.7in]{geometry}
\usepackage{graphicx}
\usepackage{flafter}
\usepackage{xcolor}
\usepackage{booktabs,tabularx,array}
\usepackage{caption}
\usepackage{titlesec}
\usepackage[numbers,sort&compress]{natbib}
\usepackage{tikz}
\usepackage{balance}
\usepackage{hyperref}

\definecolor{mmfactblue}{HTML}{183B70}
\definecolor{mmfactorange}{HTML}{F47735}
\definecolor{ink}{HTML}{181C23}
\definecolor{muted}{HTML}{606874}
\hypersetup{colorlinks=true,linkcolor=mmfactblue,citecolor=mmfactblue,
  urlcolor=mmfactblue,
  pdftitle={MMFact: Simplifying Multimodal Misinformation Detection with Shared Resources and Tools},
  pdfauthor={Yuchen Miao, Zijun Wang, Mingtai Zhang, Xiaoyi Chen, Sihan Deng, Xianglin Ding},
  pdfsubject={Shared materials and tools for fairer multimodal misinformation detection comparisons}}
\titleformat{\section}{\large\bfseries\color{ink}}{\thesection}{0.65em}{}
\titlespacing*{\section}{0pt}{11pt plus 2pt minus 1pt}{5pt}
\titleformat{\paragraph}[runin]{\bfseries}{\theparagraph}{0pt}{}[.]
\titlespacing*{\paragraph}{0pt}{6pt plus 1pt minus 1pt}{0.45em}
\newcommand{\toolbox}{MMFact}

\definecolor{badgegray}{HTML}{555555}
\definecolor{badgegithub}{HTML}{181717}
\definecolor{badgedataset}{HTML}{FFD21E}
\definecolor{badgeckpt}{HTML}{007EC6}
\newsavebox{\badgeleftbox}
\newsavebox{\badgerightbox}
\newlength{\badgeleftwidth}
\newlength{\badgetotalwidth}
\newcommand{\resourcebadge}[5]{%
  \begingroup
  \sbox{\badgeleftbox}{\color{white}\fontfamily{phv}\fontsize{8.25bp}{10bp}\selectfont
    \raisebox{-2bp}{\includegraphics[height=10.5bp]{#1}}\hspace{4bp}#2}%
  \sbox{\badgerightbox}{\color{#5}\fontfamily{phv}\fontsize{8.25bp}{10bp}\selectfont#3}%
  \setlength{\badgeleftwidth}{\dimexpr\wd\badgeleftbox+10bp\relax}%
  \setlength{\badgetotalwidth}{\dimexpr\badgeleftwidth+\wd\badgerightbox+10bp\relax}%
  \begin{tikzpicture}[baseline=-3bp]
    \path[use as bounding box] (0,0) rectangle (\badgetotalwidth,15bp);
    \begin{scope}
      \clip[rounded corners=2.25bp] (0,0) rectangle (\badgetotalwidth,15bp);
      \fill[badgegray] (0,0) rectangle (\badgeleftwidth,15bp);
      \fill[#4] (\badgeleftwidth,0) rectangle (\badgetotalwidth,15bp);
    \end{scope}
    \node[anchor=west,inner sep=0pt,text=white] at (5bp,7.5bp) {\usebox{\badgeleftbox}};
    \node[anchor=west,inner sep=0pt,text=#5,xshift=5bp] at (\badgeleftwidth,7.5bp) {\usebox{\badgerightbox}};
  \end{tikzpicture}%
  \endgroup
}

\begin{document}
\twocolumn[{
\begin{minipage}{\textwidth}
  \raggedright
  \includegraphics[width=1.5in,trim=90bp 115bp 90bp 110bp,clip]{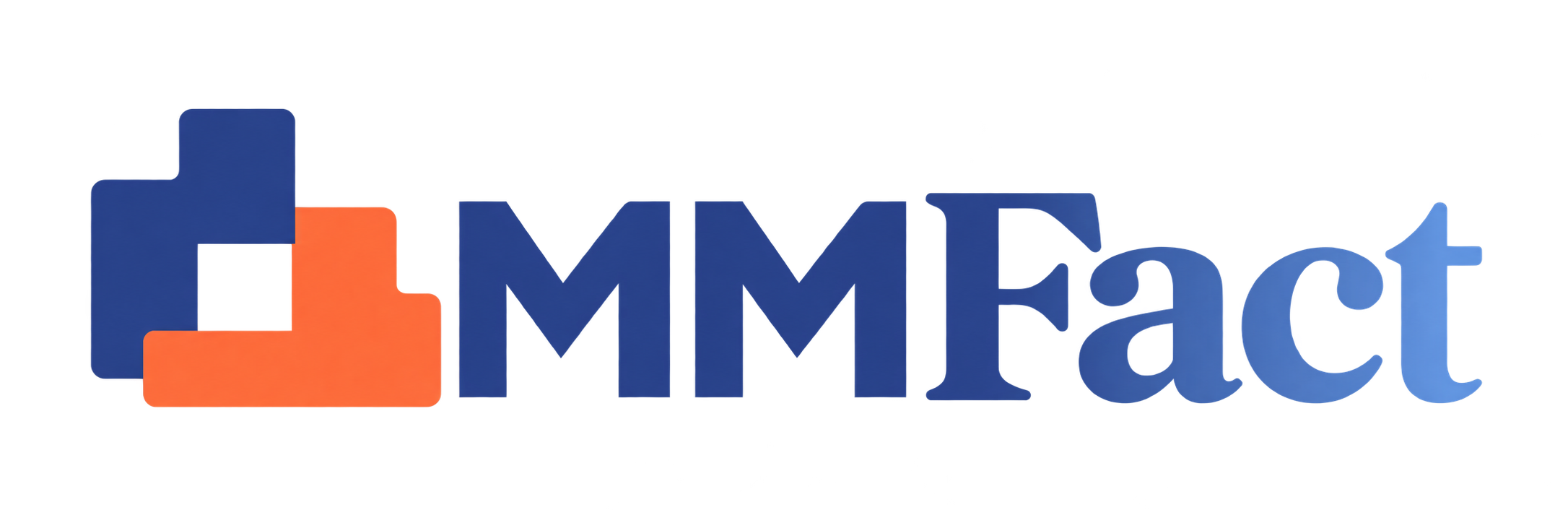}\par
  \vspace{5pt}
  {\color{mmfactblue}\rule{\textwidth}{0.6pt}}\par
  \vspace{13pt}
  {\fontsize{22}{26}\selectfont\bfseries\color{ink}
  MMFact: Simplifying Multimodal Misinformation\par
  Detection with Shared Resources and Tools\par}
  \vspace{12pt}
  {\large Yuchen Miao\textsuperscript{1,*}\quad
  Zijun Wang\textsuperscript{1,*}\quad
  Mingtai Zhang\textsuperscript{2}\quad
  Xiaoyi Chen\textsuperscript{1}\quad
  Sihan Deng\textsuperscript{3}\quad
  Xianglin Ding\textsuperscript{1}\par}
  \vspace{3pt}
  {\color{muted}\textsuperscript{1}Northeastern University\quad
  \textsuperscript{2}Guangxi University\quad
  \textsuperscript{3}Zhengzhou University\par}
  \vspace{11pt}
  \href{https://github.com/yuchenmiao04/MMFact}{\resourcebadge{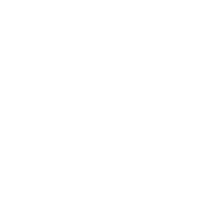}{GitHub}{Repo}{badgegithub}{white}}\hspace{5pt}%
  \href{https://huggingface.co/datasets/MMFact/MMFact}{\resourcebadge{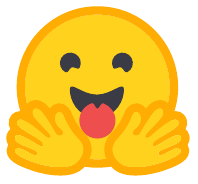}{Hugging Face}{Dataset}{badgedataset}{ink}}\hspace{5pt}%
  \href{https://huggingface.co/MMFact/MMFact-CKPT}{\resourcebadge{images/huggingface-mark.pdf}{Hugging Face}{CKPT}{badgeckpt}{white}}\par
  \vspace{17pt}
\end{minipage}
}]
\begingroup
\renewcommand{\thefootnote}{*}
\footnotetext[0]{Yuchen Miao and Zijun Wang contributed equally to this work.}
\endgroup

\section*{Abstract}
Multimodal misinformation detection methods often rely on different preprocessing pipelines and auxiliary resources, making experimental comparisons difficult to interpret. We present \toolbox, a collection of shared resources and tools that simplifies this research workflow. It brings together 47 detector implementations and baselines, three multimodal datasets, and prepared materials ranging from semantic features to OCR, image descriptions, and external evidence. Compatible methods can use the same materials, reducing variation introduced by preprocessing and supporting fairer comparisons. A modular training and evaluation framework preserves method-specific designs while providing common experimental procedures. By treating prepared materials as a shared research resource, \toolbox{} helps researchers focus on developing and evaluating detection methods.

\section{Introduction}
Multimodal misinformation detection draws on the content of news text and images as well as their relationships. A detector may require semantic representations, text recognized within images, visual regions, or external evidence. Preparing these inputs often involves separate models and services, adding substantial work before the detector itself can be studied.

These preparation choices also affect comparison. When methods use different image descriptions, feature extractors, or retrieved evidence, an observed performance difference can reflect changes in their inputs as well as their designs. Sharing compatible resources helps control this source of variation and provides a clearer basis for evaluating methods.

Toolboxes such as MMRec~\cite{mmrec} and FaKnow~\cite{faknow} illustrate the value of common infrastructure for implementing and comparing models. \toolbox{} extends this approach with a companion collection of prepared materials for multimodal misinformation detection. It combines reusable resources with configurable models, training, and evaluation. The emphasis is on supporting fairer comparisons through shared inputs while retaining the specialized information and computation that distinguish individual methods.

\section{Datasets and Reusable Materials}
The collection covers Weibo-17~\cite{attrnn}, Weibo-21~\cite{mdfend}\footnote{The original official Weibo-21 dataset was unavailable; we reconstructed this version based on the original paper.}, and GossipCop from FakeNewsNet~\cite{fakenewsnet}. Table~\ref{tab:datasets} summarizes the packaged multimodal versions, which contain 26,361 news records in total. Fixed training, validation, and test splits are shared across experiments. Weibo-17 retains EANN's released split assignments~\cite{eann} after cleaning, with one cross-split duplicate removed from the validation set. Weibo-21 uses our fixed 8:1:1 repartition. Auxiliary domain and event labels support methods that require this additional structure.

\begin{table}[ht]
\caption{The datasets prepared for \toolbox. Images count distinct image assets within each dataset.}
\label{tab:datasets}
\centering
\small
\begin{tabular*}{\columnwidth}{@{\extracolsep{\fill}}lrrrr@{}}
\toprule
Dataset & Train & Valid. & Test & Images \\
\midrule
Weibo-17~\cite{attrnn} & 5,415 & 842 & 1,465 & 7,717 \\
Weibo-21~\cite{mdfend} & 4,807 & 601 & 601 & 5,877 \\
GossipCop~\cite{fakenewsnet} & 8,842 & 1,894 & 1,894 & 12,014 \\
\bottomrule
\end{tabular*}
\end{table}

Beyond the original text and images, \toolbox{} organizes the materials in Table~\ref{tab:materials}. These resources address several recurring needs in multimodal detection, from basic content encoding to evidence-based reasoning.

\begin{table}[ht]
\caption{An overview of the reusable materials.}
\label{tab:materials}
\centering
\small
\renewcommand{\arraystretch}{1.08}
\begin{tabularx}{\columnwidth}{@{}>{\raggedright\arraybackslash}p{0.32\columnwidth}>{\raggedright\arraybackslash}X@{}}
\toprule
Resource & Representative contents \\
\midrule
Standardized data & News text and images, fixed splits, domain and event labels \\
\addlinespace[3pt]
Multimodal inputs and features & Text, image, and cross-modal representations; regions, crops, and frequency information \\
\addlinespace[3pt]
Semantic annotations & OCR, image descriptions, translations, summaries, entities, and linguistic information \\
\addlinespace[3pt]
External and generated evidence & Encyclopedia content, retrieval evidence, arguments, and multi-perspective rationales \\
\addlinespace[3pt]
Training resources & Vocabularies, initial word vectors, reference collections, and related structures \\
\bottomrule
\end{tabularx}
\end{table}

\paragraph{Representations and visual information}
Prepared text, image, and cross-modal representations support methods that work with semantic features. Region, crop, and frequency information preserve complementary visual cues required by other designs. Preparation can vary across datasets to accommodate differences in language and document length.

\paragraph{Descriptions and evidence}
OCR captures text appearing within an image, while image descriptions express its visible content. Translations, summaries, entities, and linguistic annotations provide additional textual views. Encyclopedia content and retrieval results support methods that consult external information; generated arguments and rationales support methods that reason over multiple perspectives. These materials remain auxiliary inputs, separate from the news labels.

\paragraph{Training support}
The collection also includes vocabularies, initial word vectors, and reference resources. Dataset-dependent fitting uses the training split, keeping validation and test data outside this preparation stage. These resources support subsequent model training without requiring each implementation to repeat the same preparation work.

\paragraph{Shared resources for fairer comparisons}
When two methods require the same type of compatible representation or auxiliary information, they can use the same prepared material. This reduces differences due to feature extraction, generation, or retrieval choices and helps attribute experimental differences to the methods being compared. Fixed auxiliary content also keeps repeated runs from receiving changing responses from external services.

Sharing follows the requirements of each method. Specialized region reasoning, frequency processing, and method-specific evidence retain their own inputs. Trainable encoders likewise remain trainable when required. These boundaries allow common resources to support more controlled comparisons while preserving the core designs of the detectors.

\section{Architecture and Key Features}
\begin{figure}[ht]
  \centering
  \includegraphics[width=\columnwidth]{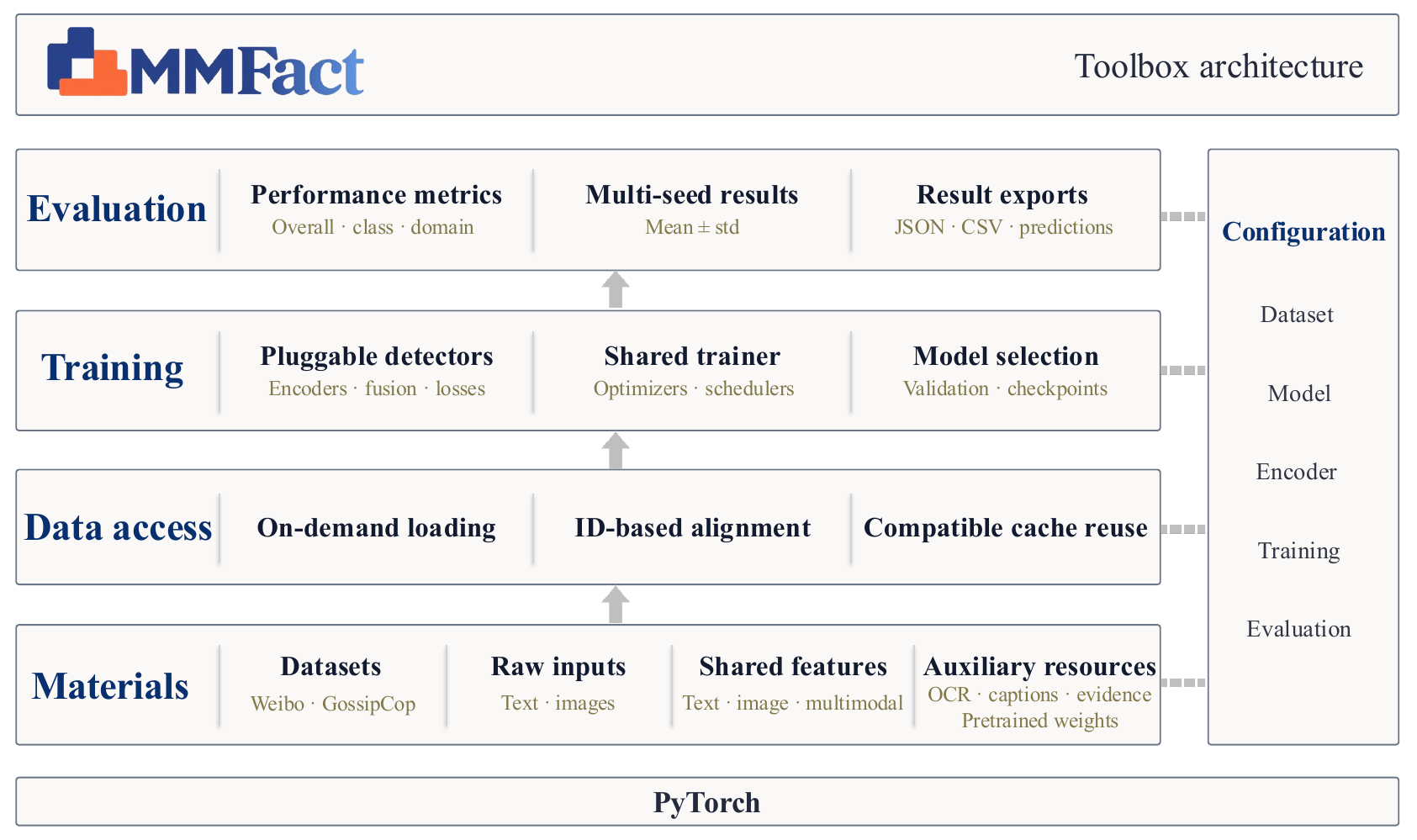}
  \caption{An overview of the \toolbox{} toolbox.}
  \label{fig:architecture}
\end{figure}

Building on this resource collection, \toolbox{} organizes data access, model training, and evaluation through a common configuration system. Figure~\ref{fig:architecture} illustrates how these components work together.

\paragraph{Data access}
Each detector receives the records, representations, and auxiliary information it needs from the shared collection. Common data access keeps these inputs aligned and allows users to obtain the required materials without downloading every resource.

\paragraph{Models and training}
The toolbox integrates 47 implementations, including eight baselines and 39 models associated with research papers. A shared training framework supports their different fusion strategies and objectives. Methods can consume prepared features or learn from textual and visual inputs, according to their designs.

\paragraph{Evaluation}
Common evaluation procedures provide overall, class-wise, and domain-wise results where applicable. They include standard classification metrics such as accuracy, precision, recall, and F1. Validation-based model selection and repeated runs with different random seeds support consistent reporting across methods.

\paragraph{Configuration}
Shared configuration controls dataset, model, encoder, and training choices within a consistent experimental framework.

\section{Conclusion}
\toolbox{} brings shared preprocessing materials and experimental tools together for multimodal misinformation detection. By combining compatible resources with common training and evaluation procedures, it supports fairer method comparisons and reduces repeated preparation work. Its modular design preserves specialized methods and provides a practical foundation for developing and evaluating new detectors.

\bibliographystyle{ACM-Reference-Format}
\bibliography{references}
\end{document}